# THE REASONS OF LINE BROADENING OF LATTICE OSCILLATIONS IN SPECTRUM SMALL FREQUENCIES OF P-CHLORONITROBENZENE

M. A. Korshunov

*LV Kirensky Institute of Physics, Siberian Branch of Russian Academy of Science, Krasnoyarsk, 660036 Russia*
*E-mail: mkor@iph.krasn.ru*

**Abstract:** Raman spectroscopies are carried out by the Method polarizable examinations of the lattice oscillations of p-chloronitrobenzene at temperature 293 K. The led matching of spectrums of p-chloronitrobenzene with p-bromochlorobenzene and p-dibromobenzene has shown that the significant line broadening in p-chloronitrobenzene is caused not only clutter in allocation of molecules concerning parasubstitution benzol. It is necessary to consider also and clutter in rotational displacements of the nitro of group concerning a plane of a molecule. Calculations of frequency spectra were led on a method the Dyne.

In a number of operations spectrums of chips consisting as from centrosymmetric, and not centrosymmetric molecules [1] were studied. Chips of builders were isomorphous. Crystallization of not centrosymmetric molecules in centrosymmetric space group is possible at statistically disorder layout of molecules concerning parasubstitution rules of benzol. On their instance it is possible to observe agency of statistical clutter on a line broadening. In the yielded operation appearance of additional lines caused is not observed by clutter.

For comparative learning isomorphous chips of p-dibromobenzene, p-bromochlorobenzene and p-chloronitrobenzene have been selected. There are their X-ray diffraction data [2]. On X-ray diffraction data p-bromochlorobenzene and p-chloronitrobenzene crystallize in the space group $P2_1/a$ with two molecules in unit cell owing to statistically disorder layout of molecules concerning parasubstitution benzol. In a spectrum of the lattice oscillations of such perfect crystals it should be observed six lines, the molecules caused by rotational oscillations.

In the yielded operation parameters of the polarized Raman spectrums of the lattice oscillations of p-dibromobenzene, p-bromochlorobenzene and p-chloronitrobenzene are studied. In patterns the polarized spectrums of studied chips are presented. As we see, spectrums are similar among themselves.

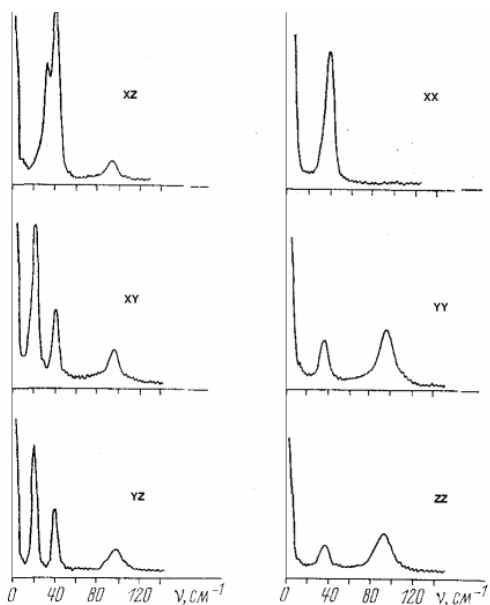

Pattern 1.
Spectrum of p-dibromobenzene.

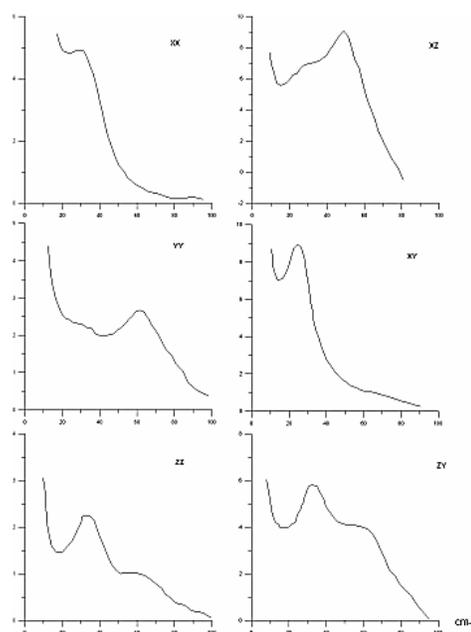

Pattern 3.
Spectrum of p-chloronitrobenzene.

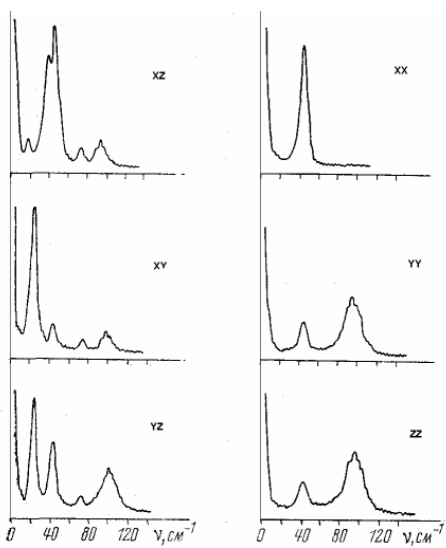

Pattern 2.
Spectrum of p-bromochlorobenzene

Outcomes of experiment are presented in the table. For p-chloronitrobenzene the improved data on matching with operation [3] are cited.

Table.
Values of frequencies ($\nu$), half width ($\delta$) and oscillation modes par substitution Benzol at temperature 293K.

| P-$C_6H_4Br_2$ | $\nu$, cm$^{-1}$ | 20.0 | 37.4 | 38.5 | 39.4 | 93.0 | 97.0 |
| | $\delta$, cm$^{-1}$ | 2.0 | 2.3 | 2.4 | 2.2 | 13.0 | 10.5 |
| P-$C_6H_4BrCl$ | $\nu$, cm$^{-1}$ | 22.5 | 41.0 | 46.0 | 42.0 | 93.0 | 100.0 |

| | δ, cm$^{-1}$ | 3.9 | 6.0 | 7.0 | 5.5 | 16.8 | 12.8 |
|---|---|---|---|---|---|---|---|
| P-C$_6$H$_4$ClNO$_2$ | ν, cm$^{-1}$ | 24.5 | 34.0 | 30.0 | 47.5 | 63.5 | 70.0 |
| | δ, cm$^{-1}$ | 12.5 | 18.5 | 20.0 | 23.0 | 26.0 | 26.0 |
| | modes | B$_g$ | A$_g$ | A$_g$ | B$_g$ | A$_g$ | B$_g$ |

In the table values of frequencies, linewidths, and also oscillation mode of six intensive lines of the molecules linked to rotational oscillations are resulted. As it is visible for statistically disorder chip of p-bromochlorobenzene of a spectral line more widely, than in the arranged p-dibromobenzene. This broadening is caused by clutter in layout of molecules of p-bromochlorobenzene. Analogous lines in p-chloronitrobenzene are much wider, that it is difficult to explain only presence of statistical clutter. This broadening is saved and at 77K.

For an explanation of experimental outcomes calculations of frequencies of the lattice oscillations of studied chips have been led. Interacting between molecules was presented on a method atom - atom of potentials [4]. Molecular composition was accepted absolutely hard. The frequency spectrum was on a method the Dyne [5]. The bar graphs gained at calculations, show probability of manifestation of spectral lines in the selected frequent interval.

From the analysis of the experimental and calculated spectrums it is discovered, that six intensive lines in a spectrum of the mixed chip are linked to orientation oscillations. Lines of p-chloronitrobenzene with frequencies of 24.5 and 34.0 cm-1 are caused by oscillations basically around of an axis with the greatest moment of inertia. Lines of 30.0 and 47.5 cm-1 with an average, and lines of 63.5 and 70.0 cm$^{-1}$ with the least moment of inertia.

As have shown scalings the value of scatter of frequencies it is linked to that for what oscillations calculation is led. The greatest scatter of frequencies is observed for lines linked with oscillations around of an axis with the greatest moment of inertia and the least with the least moment of inertia. On the average for scatter of frequencies of p-bromochlorobenzene linked to statistical clutter concerning parasubstitution halogens makes about 2-3 cm$^{-1}$. For p-chloronitrobenzene of 3-5 cm$^{-1}$ that does not explain the significant line broadening in a spectrum of small frequencies. Therefore calculations in view of not only statistical clutter of parasubstitution benzol, but also in view of statistical clutter on rotational displacements of the nitrogroup around of axis C-N have been led. The scatter of frequencies has made from 3 up to 10 cm$^{-1}$ depending on the given clutter. The value of this scatter will already agree with experimental data.

Thus, Raman spectrums of

small frequencies of p-chloronitrobenzene are similar to the analogous polarized spectrums of pure p-dibromobenzene and statistically разупорядоченного p-bromochlorobenzene that confirms X-ray diffraction data about crystallization of these substances in the same space group. In a spectrum it is discovered six intensive lines linked with orientation oscillations. The line broadening in p-bromochlorobenzene is linked to statistical clutter in layout of molecules concerning parasubstitution benzol. The registration only it condition in case of p-chloronitrobenzene does not explain the significant broadening of analogous lines. For correct interpretation it is necessary to consider and statistical clutter in rotational displacements of the nitro of group concerning axis C-N.